\documentclass[epj]{webofc}
\usepackage[varg]{txfonts}
\usepackage{mathtools}
\usepackage{orcidlink}

\hypersetup{colorlinks=true,citecolor=blue,urlcolor=blue,linkcolor=blue}
\graphicspath{{figures/}}

\newcommand{\Mc}{\mathcal M}
\newcommand{\Kc}{\mathcal K}
\newcommand{\Dc}{\mathcal D}

\newcommand{\df}{\mathrm{df}}

\newcommand{\bs}[1]{\boldsymbol{#1}}

\woctitle{MESON2026 -- 18th International Workshop on Meson Physics}
\hypersetup{
pdftitle={Meson spectroscopy from lattice QCD},
 pdfauthor={Andrew W. Jackura},
 pdfsubject={MESON2026 plenary proceedings}
 }
 
\begin{document}

\title{Meson spectroscopy from lattice QCD}

\author{\firstname{Andrew} \lastname{Jackura}\inst{1,2}\fnsep
\thanks{\email{awjackura@wm.edu}}\orcidlink{0000-0002-3249-5410}}
\institute{Department of Physics, William \& Mary, Williamsburg, VA 23187, USA
\and
Theory Center, Thomas  Jefferson  National  Accelerator  Facility, Newport  News,  Virginia  23606,  USA}

\abstract{
Determining the meson spectrum directly from Quantum Chromodynamics (QCD) remains a central challenge, since QCD is strongly coupled at low energies and most excited states appear as resonances in hadronic decay channels. 
Advances in lattice QCD and scattering theory now provide a systematic framework for extracting meson spectra and resonance properties from first principles. 
Here I present an overview of this program and review current state-of-the-art calculations in meson spectroscopy from lattice QCD. 
I highlight recent progress in the extraction of resonant and coupled-channel amplitudes, along with frontier developments in three-body dynamics and the determination of meson photo-couplings. 
Together, these advances are bringing us closer to a comprehensive first-principles description of the excited meson spectrum.
}

\maketitle

\section{Introduction}
\label{sec:introduction}

The excited hadron spectrum emerges from the strong interactions of quarks and gluons.
Most excited states are resonances, whose masses, widths, and couplings are defined by poles of scattering amplitudes.
Determining these amplitudes is therefore central to hadron spectroscopy.
Lattice QCD provides a nonperturbative approach, but its finite Euclidean spacetime yields discrete energies rather than asymptotic scattering states.
Finite-volume methods connect these energies to on-shell amplitudes, establishing a broad program of hadronic resonance studies~\cite{Luscher:1990ux,Briceno:2017max}.
We review the basics of finite-volume spectroscopy and the mapping to two- and three-hadron amplitudes, and discuss recent results.

\section{Finite-volume spectroscopy}
\label{sec:spectroscopy}

Consider a cubic spatial volume of side length $L$, with periodic boundary conditions.\footnote{Finite lattice spacing introduces an additional systematic uncertainty~\cite{Hansen:2024cai}. We neglect lattice-spacing effects in this work.}
The allowed total momenta are quantized as $\bs P=2\pi\bs d/L$, where $\bs d$ is an integer vector.
For a chosen set of quantum numbers, we construct a basis of interpolating operators $\{\mathcal O_i\}$ and evaluate the matrix of two-point functions
\begin{equation}
 C_{ij}(t)=\langle0|\mathcal O_i(t)\mathcal O_j^\dagger(0)|0\rangle_L
 =\sum_n Z_i^{(n)}Z_j^{(n)*}e^{-E_nt}.
 \label{eq:correlator}
\end{equation}
Here $Z_i^{(n)}=\langle0|\mathcal O_i|n,L\rangle$ for unit-normalized finite-volume states $|n,L\rangle$.
The operator basis must resolve the states in the energy region of interest.
Large bases combining single-hadron-like quark bilinears, including gauge-covariant derivatives, and products of momentum-projected meson operators have proved effective in determining the low-lying spectrum~\cite{Dudek:2010wm,Thomas:2011rh,Dudek:2012gj}.
The operators are projected into irreducible representations of the cubic group at rest, or the little group that preserves $\bs P$ in moving frames.
These representations generally receive contributions from several continuum angular momenta.

A particularly powerful technique to extract the finite-volume spectrum is the variational method, implemented through the generalized eigenvalue problem (GEVP)~\cite{Michael:1985ne,Luscher:1990ck,Blossier:2009kd},
\begin{equation}
 \mathbf C(t)\bs v_n(t,t_0)
 =\lambda_n(t,t_0)\mathbf C(t_0)\bs v_n(t,t_0).
 \label{eq:gevp}
\end{equation}
The principal correlators $\lambda_n(t,t_0)\sim e^{-E_n(t-t_0)}$ determine the energies, while the eigenvectors define linear combinations of operators with improved overlap with individual states.
Corrections to this exponential behavior arise from states not resolved by the finite basis.
Stability under changes of $t_0$, fit range, and operator basis is therefore important~\cite{Dudek:2010wm}.
Optimized single-meson operators can themselves be used to construct two-meson operators with reduced contamination from excited constituents~\cite{Dudek:2012gj}.

Distillation makes these large correlation matrices practical by projecting quark fields onto low-lying spatial Laplacian eigenvectors and reusing quark propagation information across operators~\cite{HadronSpectrum:2009krc}.
Varying $L$, $\bs P$, and the lattice irreps supplies additional kinematic constraints on the amplitudes.
The resulting energies are eigenvalues of QCD in a finite volume, rather than resonance masses.
The L\"uscher framework relates this discrete spectrum to infinite-volume scattering amplitudes~\cite{Luscher:1990ux,Briceno:2017max}, as illustrated by the charmonium example in Fig.~\ref{fig:charmonium}.
The amplitudes are then continued to their poles, propagating statistical, parametrization, and lattice uncertainties.
\begin{figure}[t]
 \centering
 \includegraphics[width=\textwidth]{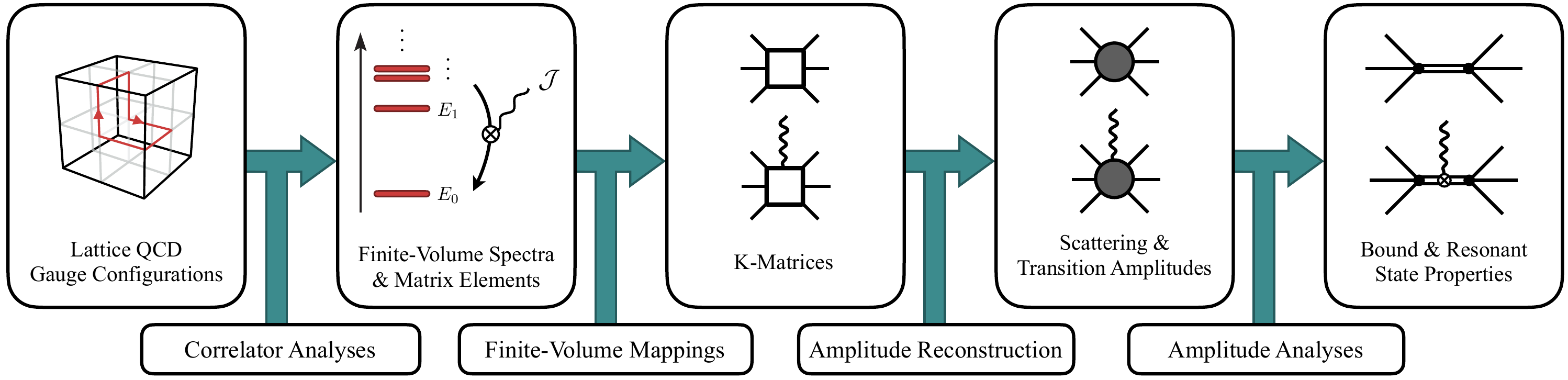}
 \caption{Workflow for extracting the spectrum and couplings of hadrons from QCD. Lattice QCD
gauge configurations are used to determine finite-volume spectra and matrix elements from multi-hadron
correlation functions. Infinite-volume K matrices are then constrained using nonperturbative finite-volume
mappings. Finally, the spectral content is extracted from the reconstructed amplitudes.}
 \label{fig:methodology}
\end{figure}
%

\section{Two-hadron systems}
\label{sec:two}

To illustrate the connection to scattering, consider an elastic partial wave of two spinless hadrons.
We choose the amplitude normalization in which
\begin{equation}
 \Mc_2(s)=\frac{1}{\Kc_2^{-1}(s)-i\rho(s)},
 \qquad \rho(s)=\frac{\xi q^\star}{8\pi\sqrt{s}},
 \label{eq:twobody}
\end{equation}
where $s=E^{\star2}$, $q^\star$ is the relative momentum in the center-of-momentum frame, and $\xi=1/2$ for identical particles and $1$ otherwise.
On the physical axis, unitarity fixes $\operatorname{Im}\Mc_2^{-1}=-\rho$.
The K matrix is real in this region, containing the remaining dynamics including crossed-channel effects.
For coupled channels, Eq.~\eqref{eq:twobody} becomes a matrix relation, with diagonal phase space and channel-dependent thresholds.

The finite-volume spectrum constrains the infinite-volume K matrix through the \emph{quantization condition}~\cite{Luscher:1986pf,Rummukainen:1995vs,Kim:2005gf,Hansen:2012tf,Briceno:2014oea},
\begin{equation}
 \det\!\left[\mathbf K_2^{-1}(E^\star)
             +\mathbf F_2(E,\bs P,L)\right]=0.
 \label{eq:qc2}
\end{equation}
The determinant runs over channels and partial waves in the chosen lattice irrep.
The known $\mathbf F_2$ characterizes the finite-volume geometry and kinematics.
Fitting the solutions of Eq.~\eqref{eq:qc2} to measured energies constrains $\mathbf K_2$, and hence the amplitude via Eq.~\eqref{eq:twobody}.
The relation neglects exponentially suppressed effects and requires all relevant on-shell channels.

Bound states and resonances are then identified by continuing the amplitude to the appropriate sheets of the complex energy plane.
Near an isolated pole we may write
\begin{equation}
 \Mc_{2,ab}(s)=\frac{g_a g_b}{s_R-s}
 +\text{terms regular at }s_R,
 \label{eq:pole}
\end{equation}
where the pole position defines the resonance mass and width, $\sqrt{s_R}=m_R-i\Gamma_R/2$.
The residues determine the channel couplings $g_a$.
Pole stability must be tested under changes of parametrization.
This methodology, the so-called L\"uscher framework, is illustrated in Fig.~\ref{fig:methodology}.

\begin{figure}[t]
 \centering
 \includegraphics[width=\textwidth]{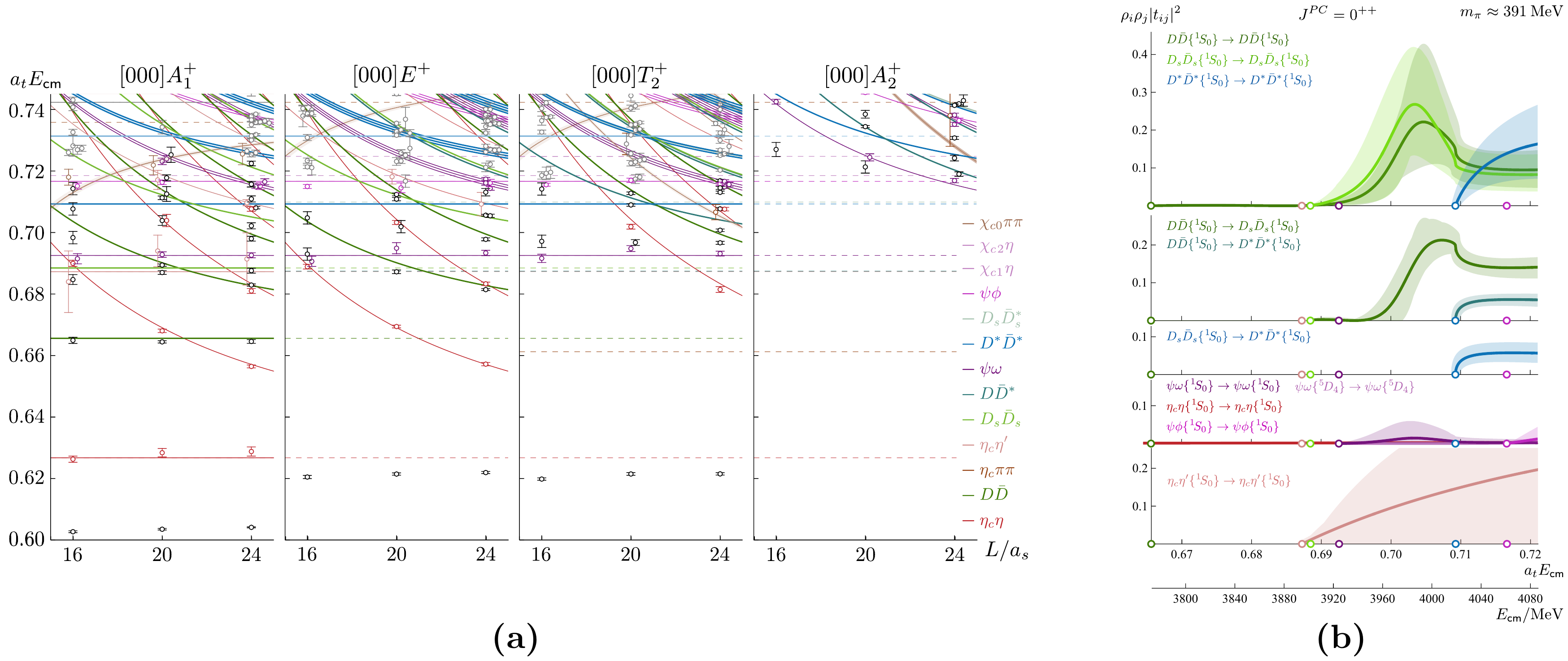}
 \caption{$J^{PC} = 0^{++}$ charmonium spectrum at $m_\pi\approx 391$ MeV.
 (a) Finite-volume energies in four positive-parity rest-frame
 irreps. Curves show noninteracting energies and dashed lines mark
 thresholds. (b) Resulting phase-space weighted squared amplitudes from the coupled-channel analysis of
 the finite-volume spectrum.
 Circles on the energy axes mark thresholds.
 Figure adapted from Ref.~\cite{Wilson:2023anv}.}
 \label{fig:charmonium}
\end{figure}
Over the past two decades there have been many applications of this framework~\cite{Padmanath:2022cvl,Whyte:2024ihh,Lyu:2023xro,Dudek:2012xn,Dudek:2014qha,Fahy:2014jxa,Wilson:2014cna,Green:2014dea,Wilson:2015dqa,Bolton:2015psa,Bulava:2015qjz,Junnarkar:2015jyf,Dudek:2016esq,Dudek:2016cru,Bulava:2016mks,Briceno:2016mjc,Moir:2016srx,Doring:2016bdr,Wilson:2016rid,Briceno:2017qmb,Andersen:2017una,Brett:2018jqw,Francis:2018qch,Brett:2018fdb,Andersen:2018mau,Hanlon:2018yfv,Woss:2019hse,Wilson:2019wfr,Bulava:2019hpz,Andersen:2019ktw,Fischer:2020yvw,Woss:2020ayi,Johnson:2020ilc,Green:2021qol,Gayer:2021xzv,Green:2021sxb,Radhakrishnan:2022ubg,Bulava:2022vpq,Green:2022rjj,Bulava:2023wrz,Draper:2023boj,Rodas:2023gma,Rodas:2023nec,BaryonScatteringBaSc:2023ori,BaryonScatteringBaSc:2023zvt,Wilson:2023anv,Wilson:2023hzu,Bulava:2023uma,Skinner:2023wwb,Bulava:2024bsi,Collins:2024sfi,Yeo:2024chk,Boyle:2024hvv,Dudek:2024roh,Boyle:2024grr,Dawid:2024dgy,Vujmilovic:2024snz,Francis:2024fwf,Erben:2025zph,Lang:2025pjq}.
Coupled-channel charmonium provides a representative example~\cite{Wilson:2023anv}.
Figure~\ref{fig:charmonium}(a) shows the rest-frame spectra from three volumes, obtained using both charmonium-like and meson--meson operators.
Figure~\ref{fig:charmonium}(b) shows amplitudes determined from the $[000]A_1^+$ spectrum; the full study also uses moving-frame levels to constrain the $0^{++}$ and $2^{++}$ amplitudes.
The enhancements in the $D\bar D$ and $D_s\bar D_s$ channels are described by a single scalar resonance near $4$ GeV, coupled significantly to $D\bar D$ and $D_s\bar D_s$, with a large coupling to the closed $D^*\bar D^*$ channel.

\section{Including left-hand cuts}
\label{sec:lhc}

A particle exchanged in the $t$ or $u$ channel produces a left-hand cut after partial-wave projection.
If a finite-volume level lies on or beyond this cut, continuing a smooth, real K-matrix parametrization through the cut does not provide the required finite-volume relation.
Even above its endpoint, a nearby exchange can limit a low-order threshold expansion.

The formalism developed in Ref.~\cite{Raposo:2025dkb} isolates one-particle exchange explicitly in the diagrammatic derivation.
The remaining short-distance kernel is constrained by the spectrum, while integral equations restore the exchange ladders and give the physical amplitude.
Reference~\cite{PitangaLachini:2026lyd} applies this explicit-exchange formalism to $T_{cc}$ tetraquark candidate in isoscalar $DD^*$ scattering at $m_\pi\approx391$ MeV, treating one-pion exchange nonperturbatively.
At this mass the $D^*$ is stable, and the exchanged pion produces a cut close to the $DD^*$ threshold.
Reanalyzing the finite-volume spectrum with exchange included gives a complex subthreshold second-sheet pole,
\begin{equation}
 \sqrt{s_{T_{cc}}}=\bigl[3854(15)-i\,55(14)/2\bigr]\ \mathrm{MeV}.
 \label{eq:tcc}
\end{equation}
Its location differs qualitatively from a real virtual-state pole obtained when the exchange cut is omitted.

A complementary approach uses an $N/D$ representation of the amplitude~\cite{Dawid:2024oey},
\begin{equation}
 \Mc_2(s)=\frac{N(s)}{D(s)},
 \label{eq:nd}
\end{equation}
in which the left- and right-hand cuts are assigned to $N$ and $D$, respectively.
The finite-volume construction constrains parametrized left-hand-cut input while preserving the unitarity cut through a dispersive denominator.
Reference~\cite{Rodas:2026zmm} applies this framework to the $H$-dibaryon channel, using 61 levels from eight ensembles at the $\mathrm{SU}(3)_F$-symmetric point, with $m_\pi\approx417$ MeV, and includes the cut from pseudoscalar-meson exchange.
Effective-range and dispersive descriptions give comparable fits to the spectrum but differ below threshold.
The $N/D$ analysis finds a shallow bound state with continuum binding energy $\Delta E=3.26(1.60)(0.04)$ MeV, where the second uncertainty reflects variation of the averaged baryon mass.

\section{Three-hadron systems}
\label{sec:three}

Three-hadron finite-volume relations have been derived using several techniques~\cite{Hansen:2014eka,Hansen:2015zga,Hammer:2017kms,Mai:2017bge,Jackura:2022gib}, see Ref.~\cite{Hansen:2019nir} for a review.
In the relativistic formulation used below, the three-particle spectrum schematically satisfies
\begin{equation}
 \det\!\left[\mathbf F_3^{-1}(E,\bs P,L\mid\mathbf K_2)
            +\mathbf K_3(E)\right]=0.
 \label{eq:qc3}
\end{equation}
The determinant is over spectator momentum, pair angular momentum, and internal quantum numbers.
The $\mathbf F_3$ function depends on two-body K matrices, kinematics, and geometry, while $\mathbf K_3$ describes the remaining short-distance dynamics.
This construction has been extended to accommodate resonant two-particle subchannels~\cite{Briceno:2018aml,Romero-Lopez:2019qrt} and all three-pion isospins~\cite{Hansen:2020zhy}, with numerical implementations tested in several irreps and moving frames for nonmaximal-isospin sectors~\cite{Alotaibi:2025pxz}.

Amplitude reconstruction requires integral equations that re-sum all pairwise scattering.
We write $\Mc_3=\Dc+\Mc_{3,\df}$, where $\Dc$ contains only pairwise rescatterings and $\Mc_{3,\df}$ is the dressed short-distance contribution.
Projection to definite $J^P$ gives coupled one-dimensional equations in spectator momentum while retaining exchange singularities explicitly~\cite{Jackura:2023qtp,Briceno:2024ehy,Jackura:2025wbw}.
Many applications have appeared in the past five years, including studies of maximal-isospin three-pion interactions~\cite{Blanton:2019vdk,Hansen:2020otl,Culver:2019vvu}, three-meson pion--kaon systems near physical quark masses~\cite{Dawid:2025doq}, and resonances such as the $a_1(1260)$~\cite{Mai:2021nul}, the $\pi(1300)$~\cite{Yan:2025mdm}, and the $\omega$~\cite{Yan:2024gwp}.

A recent calculation considers $I=2$, $J^P=1^+$ three-pion scattering at $m_\pi\approx 391$ MeV~\cite{Briceno:2025yuq}, where the isovector dipion $P$ wave contains a narrow, unstable $\rho$, and the isotensor dipion $S$ wave is weakly repulsive.
Twelve rest-frame $T_1^+$ energies from three volumes constrain three-body K matrices in the ${}^3S_1$, ${}^3D_1$, and ${}^1P_1$ channels using previously determined two-pion amplitudes from the Hadron Spectrum collaboration.
\begin{figure}[t]
 \centering
 \includegraphics[width=0.94\textwidth]{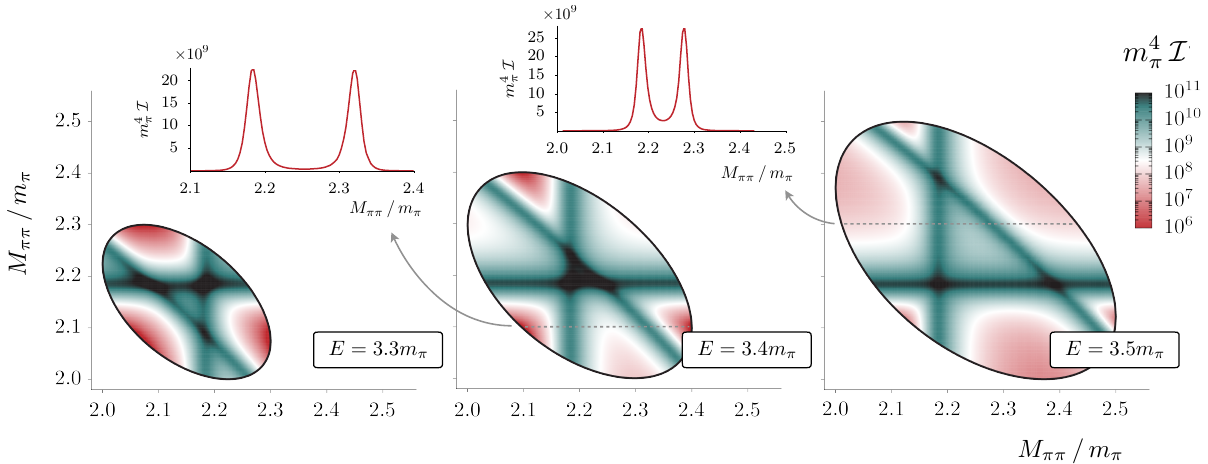}
 \caption{Dalitz distributions of isotensor three-pion $J^P=1^+$ scattering at $m_\pi\approx 391$ MeV at
 $E/m_\pi=3.3$, $3.4$, and $3.5$, using the preferred K-matrix fit and
 equal incoming/outgoing kinematics. The dipion-mass axes expose
 interfering $\rho$ bands. Figure from Ref.~\cite{Briceno:2025yuq}.}
 \label{fig:dalitz}
\end{figure}
The resulting amplitudes exhibit the dipion $\rho$ peak and a logarithmic one-pion-exchange singularity.
Averaging over acceptable K-matrix parametrizations leaves the result consistent with $\Kc_3=0$.
Summing squared moduli of symmetrized amplitudes in the constrained waveset~\footnote{A recent finite-sum expression for the exchange kernel covers arbitrary total and internal angular momenta for three massive spinless particles~\cite{Chambers:2026zfg}, facilitating larger wavesets for studying truncation effects.} gives the Dalitz distributions shown in Fig.~\ref{fig:dalitz}.
Displayed in the $J^P=1^+$ Dalitz plots are the characteristic $\rho$ bands and their interference.

\section{Photocouplings and structural observables}
\label{sec:currents}

Once pole positions and couplings are determined, we can access structural information, such as photocouplings and form factors, by probing the system with external currents.
As in the L\"uscher framework, a nonperturbative mapping relates finite-volume and infinite-volume matrix elements~\cite{Lellouch:2000pv,Jackura:2021ujw}.

The timelike pion form factor provides a direct example of extracting a production amplitude from lattice QCD.
The electromagnetic current creates a two-pion state,
\begin{equation}
 \langle\pi^+(p_1)\pi^-(p_2),\mathrm{out}|\mathcal{J}^\mu(0)|0\rangle
 =(p_1-p_2)^\mu F_\pi(s),
 \label{eq:timelike}
\end{equation}
with $s=(p_1+p_2)^2$.
Crossing relates this amplitude to the spacelike pion form factor.
In the elastic region, unitarity fixes its phase to the $\pi\pi$ $P$-wave phase shift.

Reference~\cite{Ortega-Gama:2024rqx} determines both the pion and kaon isovector form factors at $m_\pi\approx280$ MeV through the coupled $\pi\pi,K\bar K$ region.
Using the scattering matrix constrained by the spectrum, a simultaneous energy-dependent fit relates the finite-volume matrix elements to the pion and kaon production amplitudes.
\begin{figure}[t]
  \centering
  \includegraphics[width=\linewidth]{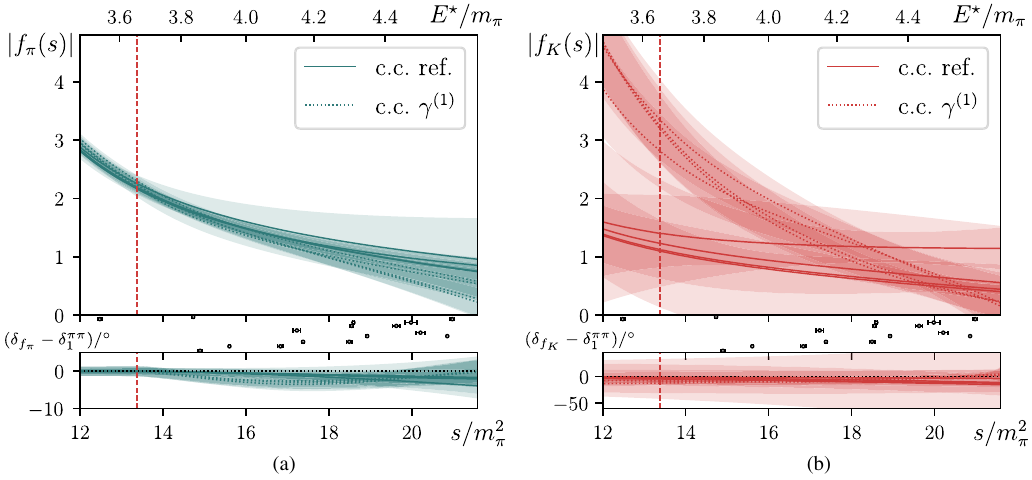}
 \caption{Timelike (a) pion and (b) kaon isovector form factors at $m_\pi\approx280$ MeV. 
 Upper panels show $|f_{\pi,K}|$, while lower panels show form-factor phases relative to the $\pi\pi$ $P$-wave phase shift.
 Solid and dotted curves use reference and alternative scattering parameterizations, with production variations and uncertainty bands. 
 Dashed lines mark the $K\bar K$ threshold. 
 Symbols between panels mark input finite-volume energies. 
 Figure taken from Ref.~\cite{Ortega-Gama:2024rqx}.}
 \label{fig:timelike}
\end{figure}
Figure~\ref{fig:timelike} shows the resulting form factors near and above the $K\bar K$ threshold.
The pion result is stable under the displayed parametrization variations, while the kaon result has larger uncertainties and greater sensitivity to the scattering parametrization, particularly near threshold.

More generally, elastic resonance form factors can be accessed through matrix elements between two-hadron states.
The $2+\mathcal{J}\to2$ framework relates finite-volume matrix elements to amplitudes with interacting initial and final states, including rescattering and single-hadron current couplings~\cite{Briceno:2015tza,Baroni:2018iau}.
Continuation to a bound-state or resonance pole in both energies defines its form factor through the double-pole residue~\cite{Briceno:2020vgp}.
A leading-order pionless lattice effective field theory test reconstructs bound-state form factors consistently across volumes, including shallow binding where direct use of finite-volume matrix elements fails~\cite{Moscoso:2026wmz}.
The first QCD implementation tests forward $\pi^+\pi^+$ vector-current matrix elements at $m_\pi\approx391$ MeV against a Ward--Takahashi prediction from the scattering amplitude~\cite{Ortega-Gama:2026hsu}.

\section{Summary and outlook}
\label{sec:outlook}

The excited hadron spectrum reflects the interplay between quark and gluon dynamics, and the many hadronic channels through which states interact and decay.
Understanding this spectrum requires a description that incorporates these features consistently.
Finite-volume methods make such a program possible within lattice QCD, connecting calculations of correlation functions to scattering amplitudes and matrix elements.
Their continued development is bringing increasingly complex hadronic systems within reach.
The long-term objective is a unified determination of hadron spectra, interactions, and electroweak properties from QCD, with uncertainties sufficiently controlled to inform the interpretation of experimental measurements.

\begin{acknowledgement}
I thank my collaborators for their contributions to our joint research.
The interpretations presented here are my own and do not necessarily reflect the original authors' views.
This work was supported by the U.S. Department of Energy (DOE) \emph{ExoHad} Topical Collaboration, Contract No.\ DE-SC0023598, and by the National Science Foundation (NSF) under Grant No. PHY-2609972.
The original articles acknowledge the resources for the individual calculations.

\end{acknowledgement}

\clearpage

\begingroup
\interlinepenalty=10000
\bibliography{references}
\endgroup

\end{document}